\documentclass[a4paper]{amsproc}

\usepackage{amssymb}
\usepackage{amscd}
\usepackage{amsfonts}
\usepackage{amssymb}
\usepackage{cite}
\usepackage{amsmath}
\usepackage{epsfig}
\usepackage{graphicx, subfigure}
\usepackage{color, graphics}
\usepackage{pdfsync}
\usepackage{hyperref}
\usepackage[all,cmtip]{xy}
\usepackage{setspace}

\theoremstyle{plain}
 \newtheorem{thm}{Theorem}
 \newtheorem{prop}{Proposition}
 
 \newtheorem{cor}{Corollary}
\theoremstyle{definition}
 \newtheorem{exm}{Example}

\theoremstyle{remark}
 \newtheorem{rem}{Remark}

\renewcommand{\le}{\leqslant}
\renewcommand{\ge}{\geqslant}
\renewcommand{\setminus}{\smallsetminus}

\allowdisplaybreaks

\newcommand{\R}{\mathbb{ R}}

\DeclareMathOperator{\pr}{pr}

\DeclareMathOperator{\diag}{diag}

\DeclareMathOperator{\Ad}{Ad}

\DeclareMathOperator{\ddim}{ddim}
\DeclareMathOperator{\dind}{dind}

\def\diag{\mathrm{diag}}

\def\px1{p_{x_1}}
\def\px2{p_{x_2}}
\def\pu1{p_{u_1}}

\title[Lax representation of homogeneous exact magnetic flows on spheres]{A Lax representation and integrability of homogeneous exact magnetic flows on spheres in all dimensions}



\author[Dragovi\'c, Gaji\'c, Jovanovi\'c]{\bfseries Vladimir Dragovi\'c, Borislav Gaji\'c, and Bo\v zidar Jovanovi\'c}

\address{
Department of Mathematical Sciences, The University of Texas at Dallas,  USA, Mathematical Institute SANU, Belgrade, Serbia}
\email{Vladimir.Dragovic@utdallas.edu}

\address{
Mathematical Institute  SANU, Belgrade, Serbia}
\email{gajab@mi.sanu.ac.rs}
\email{bozaj@mi.sanu.ac.rs}

\begin{document}

\begin{abstract} We consider motion of a material point placed in a constant homogeneous magnetic field  restricted to the sphere $S^{n-1}$.
We provide a Lax representation of the equations of motion and prove  complete integrability of those systems for any $n$.
The integrability is provided via first integrals of degree one and two.
\end{abstract}

\maketitle

\section{Introduction. The equations of motion}

Given a material point of a unit mass in a constant homogeneous magnetic field in $\R^n$
defined by the two-form
$\mathbf F=s\sum_{i<j}\kappa_{ij} d\gamma_i\wedge d\gamma_j$,
consider the motion restricted to the sphere
$S^{n-1}=\{\langle \gamma,\gamma\rangle=1\}\subset \R^n$,
where $\kappa =(\kappa_{ij})\in {so}(n)$  and $s\in \mathbb R\setminus \{0\}$.
The phase space $T^*S^{n-1}$ is a submanifold of $\mathbb{R}^{2n}(\gamma,p)$ given by the equations
$\langle\gamma,\gamma\rangle=1$, $\langle p,\gamma\rangle=0$ with the twisted symplectic form $\omega+\mathbf f$, $\omega=(dp_1\wedge d\gamma_1+\dots+dp_n\wedge d\gamma_n)\vert_{T^*S^{n-1}}$, $\mathbf f=\mathbf F\vert_{T^*S^{n-1}}$.
A motion of a material point
is described by the Hamiltonian equations on $(T^*S^{n-1},\omega+\mathbf f)$ with the Hamiltinian function $H=\frac12\langle p,p\rangle$.
In redundant variables $(\gamma,p)$, the equations are:
\begin{equation}\label{magGF}
\dot\gamma=  p,\qquad
\dot p  ={s}\kappa p+\mu\gamma, \qquad \mu={s}\langle p,\kappa\gamma \rangle-\langle p,p\rangle,
\end{equation}
where $\mu$ is the Lagrange multiplier.
From now on, we use $\ell:=[n/2]$ and consider a basis $[\mathbf e_1,\dots,\mathbf e_n]$ of $\R^n$ in which the matrix $\kappa\in so(n)$ is given by:
\begin{equation}\label{kappa}
\kappa=\kappa_{12} \mathbf e_1\wedge \mathbf e_2+\kappa_{34} \mathbf e_3\wedge \mathbf e_4 +\dots +\kappa_{2\ell-1,2\ell}\mathbf e_{2\ell-1}\wedge \mathbf e_{2\ell},
\end{equation}
where $\kappa_{2i-1,2i}\ge 0, \, i=1,\dots,\ell$.  Then equations \eqref{magGF} take the form
\begin{equation}\label{eq1}
\begin{aligned}
&{\dot\gamma}_{2i-1}=p_{2i-1},\qquad {\dot p}_{2i-1}=s\kappa_{2i-1,2i} p_{2i}+\mu\gamma_{2i-1},\\
&{\dot\gamma}_{2i} =p_{2i},\qquad\qquad  {\dot p}_{2i}=-s\kappa_{2i-1,2i} p_{2i-1}+\mu\gamma_{2i}, \qquad i=1,\dots,\ell,
\end{aligned}
\end{equation}
for $n$ even,
and, for $n$  odd, there is an additional couple of equations:
$\dot\gamma_n=p_n$ and ${\dot p}_n=\mu\gamma_n$.

These magnetic systems were obtained in \cite{DGJ2023} as a reduction of the nonholonomic problem of rolling of a ball with the gyroscope without slipping and twisting over a plane and over a sphere in $\R^n$. Also, for $n=3$ and $n=4$, we performed explicit integrations of the equations of motion in elliptic functions \cite{DGJ2023}. From our recent paper \cite{DGJ2025}, we know {the following gauge Noether integrals (see e.g.  \cite{CS1981}) of the magnetic flows:
\begin{equation}\label{linearni}
\Phi_{2i-1,2i}=\gamma_{2i-1}p_{2i}-\gamma_{2i}p_{2i-1}+ s \frac{\kappa_{2i-1,2i}}{2}\big( \gamma_{2i-1}^2+\gamma_{2i}^2\big),
\qquad i=1,\dots,\ell,
\end{equation}
and, for $\kappa_{2i-1,2i}=\kappa_{2j-1,2j}$,\footnote{
The integrals $\Psi^1_{i,j}$ and $\Psi^2_{i,j}$ are equal to
 $\Psi^1_{2i-1,2i;2j-1,2j}$ and $\Psi^2_{2i-1,2i;2j-1,2j}$ from \cite{DGJ2025}, respectively.}
\begin{equation}\label{psi}
\begin{aligned}
&\Psi_{i,j}^1=(\gamma_{2i}p_{2j-1}-\gamma_{2j-1}p_{2i})-(\gamma_{2i-1}p_{2j}-\gamma_{2j}p_{2i-1})
-s\kappa_{2i-1,2i}(\gamma_{2i-1}\gamma_{2j-1}+\gamma_{2i}\gamma_{2j}), \\
&\Psi_{i,j}^2=(\gamma_{2i-1}p_{2j-1}-\gamma_{2j-1}p_{2i-1})+(\gamma_{2i}p_{2j}-\gamma_{2j}p_{2i})
-s\kappa_{2i-1,2i}(\gamma_{2i-1}\gamma_{2j}-\gamma_{2i}\gamma_{2j-1}).
\end{aligned}
\end{equation}

In \cite{DGJ2025} we constructed one additional first integral,
$J=s^2\sum_{i=1}^{\ell}\kappa_{2i-1,2i}^2 (p_{2i-1}^2+p_{2i}^2)-\mu^2$, and proved complete integrability of the magnetic flows for any $n$ and $\kappa$ when a system allows a reduction to the cases with $n\le 6$.  We also conjectured that  magnetic systems are completely integrable for all $n$ (see \cite{DGJ2025}).

In the present note, we provide Lax representations of the equations of motion, analogous to the Lax matrix for the Neumann system given by Moser \cite{M1}.   We prove complete integrability of those systems for any $n$ and $\kappa$.
Independently, such integrability has been shown by Bolsinov, Konyaev, and Matveev in \cite{BKM2025}.
 The approach of \cite{BKM2025}, though different from ours, also has the Neumann system in the background.

The Hamiltonian formalism for magnetic geodesics in a general setting was introduced by Novikov  \cite{N1982}. Integrability
 of magnetic flows was studied in e.g. \cite{AS2020, BK2017, BJ2008, T2016, MSY2008, S2002}.

\section{Lax representations and Liouville integrability}

Here we use the full magnetic momentum maps of the $SO(n)$ and $U(\ell)$ actions on $(T^*S^{n-1},\omega+\mathbf f)$.

\begin{prop} The action $(\gamma,p)\mapsto (R\gamma,Rp)$, $R\in SO(n)$,  is Hamiltonian with the magnetic momentum map
$\Phi_s^{so(n)}\colon T^*S^{n-1}\to so(n)\cong so^*(n)$ given by
\begin{equation}\label{momPr1}
\Phi_s^{so(n)}=\gamma\wedge p+\frac{s}{2}(\kappa\gamma\otimes \gamma +\gamma\otimes \gamma \kappa).
\end{equation}
\end{prop}

For $n=2\ell$ it is also convenient to use a complex notation.
We set
$z_i=\gamma_{2i-1}+\sqrt{-1}\gamma_{2i}$, $w_i=p_{2i-1}+\sqrt{-1} p_{2i}$.
Then the equations \eqref{eq1} take the form:
\begin{align}
\label{com1}
\dot z=w, \qquad \dot w=-\sqrt{-1}sKw+\mu z, \qquad \mu={\sqrt{-1}}\frac{s}2\big(\langle Kw,\bar z\rangle- \langle z,K\bar w\rangle \big)-\langle w,\bar w\rangle,
\end{align}
where $z=(z_1,...,z_{\ell})$, $w=(w_1,...,w_{\ell})$ and
\begin{equation}\label{KK}
K=\diag(\kappa_{1,2},...,\kappa_{2\ell-1,2\ell}).
\end{equation}
In the complex notation:
$
T^*S^{2\ell-1}=\{(z,w)\in\mathbb C^{2\ell}\,\vert\, \langle \bar{z},z\rangle=1, \langle \bar{w},z\rangle+\langle w,\bar{z}\rangle=0\}.
$

\begin{prop} The action $(z,w)\mapsto (Sz,Sw)$, $S\in U(\ell)$  is Hamiltonian with the magnetic momentum map
$\Phi_s^{u(\ell)}\colon T^*S^{2\ell-1}\to u(\ell)\cong u^*(\ell)$ given by
\begin{equation}\label{momPr2}
\Phi_s^{u(\ell)}=\frac{1}{2}(w\otimes \bar z - z\otimes \bar w)+\sqrt{-1}\frac{s}{4}(Kz\otimes \bar z +z\otimes \bar zK).
\end{equation}
\end{prop}

\begin{prop}\label{pr:phis} The time derivative of the magnetic momentum maps $\Phi_s^{so(n)}$ and  $\Phi_s^{u(\ell)}$ along the equations \eqref{magGF} and \eqref{com1}
are respectively given by:
\begin{equation*}\label{eq:f0}
\dot \Phi_s^{so(n)}=\frac{s}2[\kappa,\Phi_0^{so(n)}], \qquad \dot \Phi_s^{u(\ell)}=\sqrt{-1}\frac{s}{2}[\Phi_0^{u(\ell)}, K].
\end{equation*}
\end{prop}

\begin{cor}[The Noether integrals]\label{posledica} Let $so(n)_\kappa=\{\xi\in so(n)|[\xi,\kappa]=0\}$ and $u(\ell)_K=\{\xi\in u(\ell)|[\xi,K]=0\}$ be the isotropy subalgebras of $\kappa$ and $K$ within $so(n)$ and $u(\ell)$, respectively.
Then $\pr_{so(n)_\kappa}\Phi_s^{so(n)}$ and $\pr_{u(\ell)_K}\Phi_s^{u(\ell)}$ are first integrals of the equations of motion \eqref{magGF} and \eqref{com1}, where the projection is considered with respect to $\Ad$-invariant scalar products
on $so(n)$ and $u(\ell)$.
\end{cor}

For $\kappa$ given by \eqref{kappa},  Corollary \ref{posledica} gives that $\Phi_{2i-1, 2i}$  have
geometric interpretation as components of the magnetic momentum maps:
$(\Phi_s^{so(n)})_{2i-1,2i}=\Phi_{2i-1, 2i}$, $(\Phi_s^{u(\ell)})_{i,i}=\sqrt{-1}\Phi_{2i-1, 2i}$, $i=1,\dots,\ell$.
 The $(i,j)$-th component of $\Phi_s^{u(\ell)}$,
\begin{equation*}\label{eq:psiovi}
(\Phi_s^{u(\ell)})_{i,j}=\frac12(w_i\bar z_j-z_i\bar w_j)+\sqrt{-1}\frac{s}{4}(\kappa_{2i-1,2i}+\kappa_{2j-1,2j})z_i\bar z_j,
\end{equation*}
is also a first integral for $\kappa_{2i-1,2i}=\kappa_{2j-1,2j}$.
Then the imaginary and real parts of $(\Phi_s^{u(\ell)})_{i,j}$ provide first integrals, which, multiplied by $-1/2$, coincide with
the first integrals $\Psi^1_{i,j}$ and $\Psi^2_{i,j}$ given by \eqref{psi}. Equivalently, for $\kappa_{2i-1,2i}=\kappa_{2j-1,2j}$
we have $[\kappa,\mathbf e_{2i-1}\wedge\mathbf e_{2j}\pm \mathbf e_{2i}\wedge\mathbf e_{2j-1}]=0$, and we get
integrals \eqref{psi} from the corresponding components of the magnetic momentum map $\Phi^{so(n)}_s$.

Further, in the case that there are more than one parameter $\kappa_{2i-1,2i}$ equal to zero for even $n$, or at least  parameter equal to zero for odd $n$, then there are additional
Noether integrals formed by all the components of $\Phi^{so(n)}_s$ that coincide to the components of the standard, non-magnetic momentum map $\Phi^{so(n)}_0$:
\begin{equation}\label{so(r)}
\Phi_{k,j}=(\Phi^{so(n)}_s)_{k,j}=\gamma_{k}p_{j}-\gamma_{j}p_{k}.
\end{equation}
%

\begin{prop} The equations  \eqref{magGF} and \eqref{com1} imply, respectively:
\[
\begin{aligned}
&\dot \Phi_s^{so(n)}=\frac{s}2[\kappa,\Phi_s^{so(n)}]+\frac{s^2}4[\gamma\otimes\gamma,\kappa^2], \quad\qquad (\gamma\otimes \gamma)^{\cdot}=[\gamma\otimes \gamma,\Phi_s^{so(n)}]+\frac{s}2[\kappa,\gamma\otimes\gamma], \\
&\dot \Phi_s^{u(\ell)}=\sqrt{-1}\frac{s}{2}[\Phi_s^{u(\ell)}, K]+\frac{s^2}{8}[z\otimes \bar z,K^2], \quad (z\otimes \bar z)^\cdot=2[\Phi_s^{u(\ell)}, z\otimes \bar z]+\frac{\sqrt{-1}s}{2}[z\otimes\bar z,K].
\end{aligned}
\]
\end{prop}

\begin{thm}\label{th:LA} Let $\Phi_s^{so(n)}$, $K$, $\Phi_s^{u(\ell)}$,  be given by \eqref{momPr1}, \eqref{KK}, and \eqref{momPr2}. Consider the matrices
\begin{equation}\label{eq:LA*}
\mathcal L(\lambda)=\lambda^2\frac{s^2}{4}\kappa^2+\lambda \Phi_s^{so(n)} +\gamma\otimes \gamma;
\quad \mathcal A(\lambda)=-\frac{s}{2}\kappa-\lambda^{-1}\gamma\otimes \gamma,
\end{equation}
and
\begin{equation}\label{eq:LA}
L(\lambda)=-\lambda^2\frac{s^2}{16}K^2+\lambda \Phi_s^{u(\ell)} +z\otimes \bar z;
\quad A(\lambda)=\sqrt{-1}\frac{s}{2}K+\lambda^{-1}2z\otimes \bar z.
\end{equation}
The equations of motion \eqref{magGF} and \eqref{com1} imply, respectively, the Lax representations
\begin{equation*}\label{eq:Lax}
\dot{\mathcal L}(\lambda)=[\mathcal L(\lambda),\mathcal A(\lambda)] \quad \text{and} \quad \dot L(\lambda)=[L(\lambda), A(\lambda)].
\end{equation*}
\end{thm}

The matrices $\mathcal L(\lambda)$ and $L(\lambda)$ in \eqref{eq:LA*} and \eqref{eq:LA} are analogous to the Lax matrix for the Neumann system  from \cite{M1}.
They are related to the symmetric pair decompositions
\[
gl(n,\mathbb R)=so(n)\oplus \{\text{symmetric matrices}\} \quad \text{and} \quad gl(\ell,\mathbb C)=u(\ell)\oplus \{\text{Hermitean matrices}\}.
\]

Set $A=s^2K^2/16=\diag(a_1,\dots,a_\ell)$.
As in the Neumann case \cite{M1}, starting from the matrix $L(\lambda)$ in \eqref{eq:LA}, we get:

\begin{thm}
The quadratic in momenta functions
\[
G_\lambda(z,w)=\sum_{1\le i<j\le \ell}\frac{\vert (\Phi_s^{u(\ell)})_{i,j}\vert^2}{(\lambda-a_i)(\lambda-a_j)}+\sum_{k=1}^\ell \frac{\vert z_k\vert^2}{\lambda-a_k}
\]
are  first integrals of the system \eqref{com1} for all $\lambda\ne a_i$, which are in involution among themselves and with
the Noether integrals, the components of $\pr_{u(\ell)_K}\Phi_s^{u(\ell)}$ and $\pr_{so(2\ell)_\kappa}\Phi_s^{so(2\ell)}$.
\end{thm}

In the case $n=2\ell-1$, we set $\kappa_{2\ell-1,2\ell}=0$ and note that the manifold
$\{\gamma_{2l}=0,p_{2l}=0\}$ is invariant under the flow of \eqref{com1}, providing a set of commuting first integrals
$G_\lambda\vert_{\gamma_{2l}=p_{2l}=0}$.

If all $\kappa_{2i-1, 2i}$ are distinct, then there are $\ell$ commuting first integrals
\begin{equation}\label{kvadratni}
F_i(z,w)=\lim_{\lambda\to a_i}(\lambda-a_i)\cdot G_\lambda=\sum_{k\ne i}\frac{\vert (\Phi_s^{u(\ell)})_{i,k}\vert^2}{a_i-a_k}+{\vert z_i\vert^2}, \qquad i=1,\dots,\ell,
\end{equation}
satisfying the relation $F_1+\dots+F_\ell=1$.
Among the Poisson commuting functions \eqref{linearni}, \eqref{kvadratni},  there are  $2\ell-1$ independent ones on $T^*S^{2\ell-1}$ and
$2\ell-2$ independent ones on $T^*S^{2\ell-2}$ (for $\kappa_{2l-1,2l}=0$, $\gamma_{2\ell}=p_{2\ell}=0$).  In terms of these first integrals, the Hamiltonian $H$ can be expressed as
\[
H=\sum_{i=1}^{\ell}\Big(\frac{s^2}{8}\kappa_{2i-1,2i}^2F_i+
\Phi_{2i-1,2i}^2-\frac{s}{2}\kappa_{2i-1,2i}\Phi_{2i-1,2i}\Big)-\frac12(\sum_{i=1}^{\ell}\Phi_{2i-1,2i})^2.
\]

\begin{thm}\label{th:intalln}
Assume that all parameters $\kappa_{2i-1, 2i}$ are distinct  and, for odd $n$, different from zero.
The magnetic  flows are Liouville integrable on $T^*S^{n-1}$ for all $n$ by means of the linear Noether integrals $\Phi_{2i-1,2i}$ and the quadratic first integrals $F_i$ obtained from the Lax representation.
\end{thm}

\section{Non-commutative integrability}

When some of $\kappa_{2i-1, 2i}$ are equal,
by adding all the Noether first integrals to $G_\lambda$, $\Phi_{2i-1,2i}$, we get non-commutative integrability.
Firstly, we consider the case $n=2\ell$.

\begin{exm}\label{primer}
Assume that we have only one pair of equal parameters, say $\kappa_{12}=\kappa_{34}$.
Then $F_1$ and $F_2$ are not defined, but we can consider the limit $F_1+F_2$ as $a_1$ tends to $ a_2$:
\[
\hat F_{12}=\sum_{k\ne 1,2}\frac{\vert (\Phi_s^{u(\ell)})_{1,k}\vert^2+\vert (\Phi_s^{u(\ell)})_{2,k}\vert^2}{a_1-a_k}+{\vert z_1\vert^2}+{\vert z_2\vert^2} \qquad (\hat F_{12}+F_3+\dots F_\ell=1).
\]

For $\kappa_{12}=\kappa_{34}\ne 0$, the first integrals $\Phi_{12},\Phi_{34},\Psi^1_{12},\Psi^2_{12}$ form a Lie algebra isomorphic to $u(2)$ with respect to the Poisson brackets with two independent $u(2)$--Casimirs
$I_1=\Phi_{12}+\Phi_{34}$ and $I_2=2(\Phi_{12})^2+2(\Phi_{34})^2+(\Psi^1_{12})^2+(\Psi^2_{12})^2$ (see \cite{DGJ2025}).
As a result, we get the algebra of first integrals $\mathcal F$ generated by
\begin{equation*}
\hat F_{12}, F_3,\dots, F_\ell,\Phi_{12},\Phi_{34},\Psi^1_{12},\Psi^2_{12},\Phi_{56},\dots,\Phi_{2\ell-1,2\ell}
\end{equation*}
with the set of functions
$\hat F_{12}, F_3,\dots, F_\ell,I_1,I_2,\Phi_{56},\dots,\Phi_{2\ell-1,2\ell}$
that commute with all the first integrals $\mathcal F$. We have $\ddim\mathcal F=2\ell$, $\dind\mathcal F=2\ell-2$, and
$\ddim\mathcal F+\dind\mathcal F=\dim T^*S^{2\ell-1}$. Therefore,  the system \eqref{com1} is completely integrable in the
non-commutative sense and the dimension of generic invariant isotropic tori is $2\ell-2$ (see \cite{MF, N, BJ2003}).

On the other hand, for $\kappa_{12}=\kappa_{34}=0$, we have the Noether integrals $\Phi_{i,j}$, $1\le i<j\le 4$.
(see \eqref{so(r)}), which form a Lie algebra isomorphic to $so(4)$ with respect to the Poisson brackets. Now, we get the algebra of first integrals $\mathcal F$ generated by
\begin{equation*}
\hat F_{12}, F_3,\dots, F_\ell,\Phi_{12}, \Phi_{13}, \Phi_{14}, \Phi_{23}, \Phi_{24}, \Phi_{34},\Phi_{56},\dots,\Phi_{2\ell-1,2\ell}
\end{equation*}
with the set of functions
$\hat F_{12}, F_3,\dots, F_\ell,I,\Phi_{56},\dots,\Phi_{2\ell-1,2\ell}$
($I=\Phi_{12}^2+\Phi_{13}^2+ \Phi_{14}^2+ \Phi_{23}^2+\Phi_{24}^2+\Phi_{34}^2$)
that commute with all first integrals $\mathcal F$. We have $\ddim\mathcal F=2\ell+1$, $\dind\mathcal F=2\ell-3$, and
we thus prove the non-commutative integrability of the system with the dimension of generic invariant isotropic tori equal to $2\ell-3$ \cite{BJ2003}.
The last statement can be obtained also from the $SO(4)$-symmetry reduction described below.
\end{exm}

To determine the dimension of the invariant isotropic tori with multiple equalities of parameters, we use the reduction procedure described in the last section of \cite{DGJ2025}. Without losing a generality, we assume
\begin{equation}\label{parametri}
\begin{aligned}
&a_1=\dots=a_{r_1}=\alpha_1, \dots, a_{r_1+\dots+r_{\rho-1}+1}=\dots=a_{r_1+\dots+r_{\rho}}=\alpha_\rho, \quad \alpha_1>\dots>\alpha_\rho>0,\\
& a_{r_1+\dots+r_{\rho}+1}=\dots=a_{2\ell}=\alpha_{\rho+1}=0,  \qquad \ell=r_1+r_2+\dots+r_{\rho}+r_{\rho+1},
\end{aligned}
\end{equation}
where, as above, $a_i=\kappa_{2i-1,2i}^2/16$. It is allowed that $r_{\rho+1}$ could be equal to zero. We have the following three rules:

\begin{itemize}

\item{} $U(r_i)$-symmetry reduction.  Whenever we have $r_i>2$ parameters among $a_i$ are equal to $\alpha_i>0$, for an arbitrary solution $(z(t),w(t))$
of the system \eqref{com1}, there exist
$S_i\in U(r_i)$, such that $S_i\cdot (z(t),w(t))$ is the solution of the corresponding problem on the sphere $S^{2\ell-1-2(r_i-2)}\subset S^{2\ell-1}$ with only
two parameters equal to $\alpha_i$ (see \cite{DGJ2025}). Thus, the dimension of the invariant tori is the same as for the magnetic flow on the sphere $S^{2\ell-1-2(r_i-2)}$ with only two parameters equal to $\alpha_i$.

\item{} $SO(2r_{\rho+1})$-symmetry reduction.
If $r_{\rho+1}>1$ parameters among $a_i$ are equal to zero, for an arbitrary solution $(z(t),w(t))$
of the system \eqref{com1}, there exist $R_{\rho+1}\in SO(2r_{\rho+1})$,
such that $R_{\rho+1}\cdot (z(t),w(t))$ is the solution of the corresponding problem on the sphere
$S^{2\ell-1-2(r_{\rho+1}-1)}\subset S^{2\ell-1}$ with only one parameter equal to zero (see \cite{DGJ2025}).

\item{} Assume $r_i\in\{1,2\}$, $i=1,\dots,\rho$, and that only one parameter could eventually be equal to zero ($r_{\rho+1}\in\{0,1\}$). For any pair of equal parameters, due to the first integrals \eqref{psi}, the dimension of the invariant tori drops by one (see Example \ref{primer}).
    Thus, the dimension of generic invariant isotropic tori is equal to
    $
    \delta=2\ell-1-\big( (r_1-1)+\dots+(r_\rho+1)\big)=(r_1+1)+\dots+(r_\rho+1)+2r_{\rho+1}-1.
    $
\end{itemize}

By applying the above rules (as an illustration, see Fig. \ref{redukcija}), we get:

\begin{thm}
Assume $n=2\ell$ and that parameters $a_i=\kappa_{2i-1,2i}^2/16$ satisfy the relations \eqref{parametri}. The magnetic geodesic flow \eqref{com1}
is completely integrable in the non-commutative sense. The dimension of generic invariant isotropic tori is
\[
\delta(S^{2\ell-1};r_1,\dots,r_{\rho},r_{\rho+1})=f(r_1)+\dots+f(r_\rho)+g(r_{\rho+1})-1,
\]
where
$f(r_i)=3$ for $r_i\ge 2$ , $f(r_i)=2$ for $r_i=1$, $i=1,\dots,\rho$, and $g(r_{\rho+1})=2$ for $r_{\rho+1}\ge 1$, and $g(r_{\rho+1})=0$ for $r_{\delta+1}=0$.
In particular, if all $\kappa_{2i-1,2i}$ are equal, then $\delta(S^{2\ell-1};\ell)=2$.
\end{thm}

\begin{figure}[h]
{\centering
{\includegraphics[width=10cm]{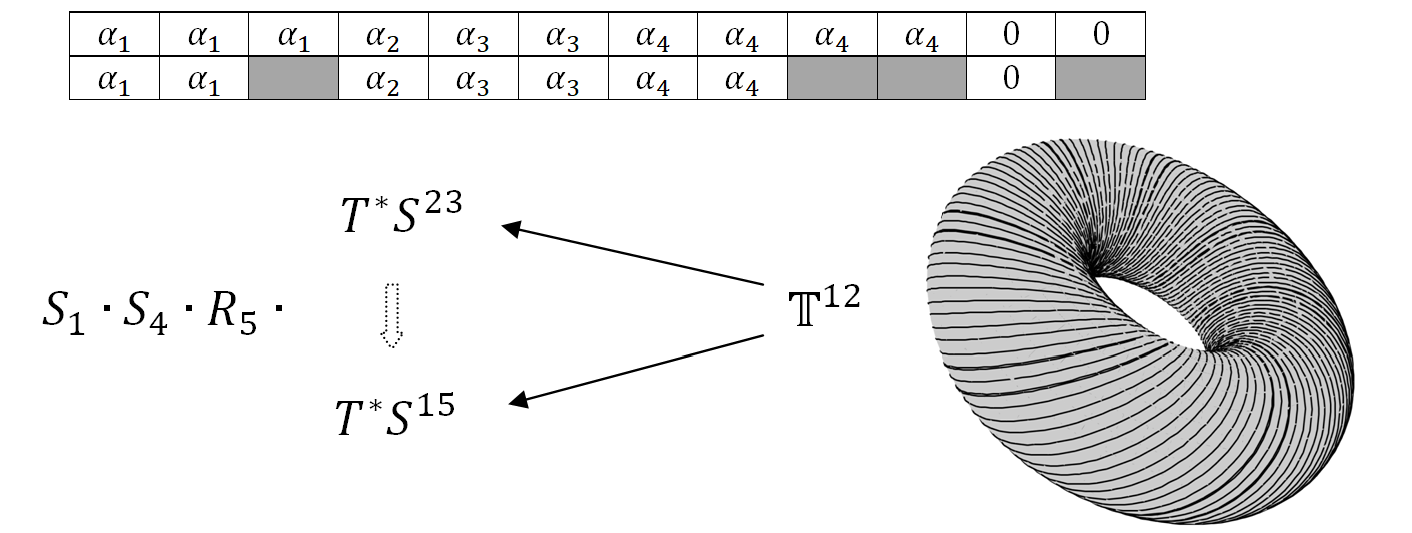}}
\caption{$\ell=12$, $\rho=4$, $r_1=3$, $r_2=1$, $r_3=2$, $r_4=4$, $r_5=2$. For every solution $(z(t),w(t))\in T^*S^{23}$, there exist rotations $S_1\in U(3), S_4\in U(4), R\in SO(4)$, such that
the solution $S_1\cdot S_4\cdot R\cdot (z(t),w(t))$ belongs to the invariant subspace $T^* S^{15}= \{z_i=w_i=0, i=3,9,10,12\}$.} \label{redukcija}
}
\end{figure}

Let $\mathcal I_{\alpha_j}$ denote the set of indices $k\in\{1,\dots,\ell\}$, such that $a_k=\alpha_j$ ($j=1,\dots,\rho+1$).
Like in Example \ref{primer}, we consider the limits $\sum_{j\in\mathcal I_{\alpha_i}} F_j$ as $a_j$ tends to $\alpha_i$, $j\in I_{\alpha_i}$ and obtain commuting first integrals
\[
\hat F_{\alpha_i}=\sum_{\alpha_k\ne \alpha_i}\frac{\sum_{p\in \mathcal I_{\alpha_i}, q\in\mathcal I_{\alpha_k}}\vert (\Phi_s^{u(\ell)})_{p,q}\vert^2}{\alpha_i-\alpha_k}+
\sum_{s\in \mathcal I_i}{\vert z_s\vert^2} \qquad (\hat F_{\alpha_1}+\dots +\hat F_{\alpha_\rho}+\hat F_{0}=1),
\]
where $\hat F_{0}\equiv 0$ if $\mathcal I_{0}=\emptyset$ ($r_{\rho+1}=0$).
The complete algebra of first integrals $\mathcal F$ is generated by
\begin{align}
&\label{komutativni} \hat F_{\alpha_1}, \dots, \hat F_{\alpha_\rho}, \hat F_{0}, \Phi_{12},\Phi_{34},\dots,\Phi_{2\ell-1,2\ell}, \\
&\label{nekom1} \Psi^1_{j,k}, \Psi^2_{j,k}, \qquad j<k, \, j,k\in \mathcal I_{\alpha_i} \qquad \text{for} \qquad r_i\ge 2, \, i=1,\dots,\rho,\\
&\label{nekom2} \Phi_{2j-1,2k-1}, \, \Phi_{2j-1,2k}, \, \Phi_{2j,2k-1}, \, \Phi_{2j,2k}, \qquad j<k, \, j,k \in \mathcal I_{0} \qquad \text{for} \qquad  r_{\rho+1}\ge 2.
\end{align}

\begin{rem}[Commutative integrability]
The magnetic systems \eqref{com1} are also Liouville integrable, where a Lagrangain toric foliation is not unique (see \cite{BJ2003}).
One set of involutive first integrals consists of first integrals \eqref{komutativni} along with the  integrals
\[
J_{\alpha_i,k}=\sum_{r_1+\dots+r_{i-1} < p < q \le k} (\Psi^1_{i,j})^2+(\Psi^2_{i,j})^2, \qquad k=r_1+\dots+r_{i-1}+2,\dots,r_1+\dots+r_{i}
\]
related to the  filtration ${u}(1)<{u}(2)<\dots<{u}(r_i)$ for every $U(r_i)$-symmetry
block with $r_i\ge 2$, $i=1,\dots, \rho+1$ (see \cite{DGJ2025}).  This includes, for $r_{\rho+1}\ge 2$, the last block, where the non-commutative integrability
follows from $SO(2r_{\rho+1})$--symmetry and the integrals \eqref{nekom2}.
In total, for $r_{\rho+1}\ge 1$, we obtain
$(\rho+1+\ell)+(r_1-1)+\dots+(r_{\rho+1}-1)=2\ell$
functions in involution satisfying the relation $\hat F_{\alpha_1}+\dots+\hat F_{\alpha_{\rho}}+\hat F_0=0$. Similarly, for $r_{\rho+1}=0$, there are $2\ell-1$
independent commuting first integrals as well. Recall that $\hat F_{0}\equiv 0$ for $r_{\rho+1}=0$.
\end{rem}

Finally, we consider the case $n=2\ell-1$ by taking $\kappa_{2\ell-1,2\ell}=0$ and $\gamma_{2\ell}=p_{2\ell}=0$ in the equation \eqref{com1}.
Again, we assume the relations \eqref{parametri}, where now $r_{\rho+1}\ge 1$. We have:

\begin{itemize}
\item{} If $\kappa_{2\ell-1,2\ell}$ is the only parameter equal to zero, the dimension of invariant tori
drops by one: $\delta(S^{2\ell-2})=\delta(S^{2\ell-1})-1$.
\item{} For $r_{\rho+1}\ge 2$, for an arbitrary motion $(z(t),w(t))$,
we can apply a rotation $R_{\rho+1}\in SO(2r_{\rho+1})$, such that $R_{\rho+1}\cdot (z(t),w(t))\in T^*S^{2\ell-2}=\{(z,w)\in T^*S^{2\ell-1}\,\vert \gamma_{2\ell}=p_{2\ell}=0\}$.
Thus, the dimension remains the same: $\delta(S^{2\ell-2})=\delta(S^{2\ell-1})$.
\end{itemize}

\subsection*{Acknowledgements}

We thank A. V. Bolsinov, A. Yu. Konaev, and V. S. Matveev for sharing with us their unpublished manuscript \cite{BKM2025}.  The research was supported by the Serbian Ministry of Science and the Simons Foundation grant no. 854861.

\end{document}